\documentclass[12pt]{article}

\usepackage[utf8]{inputenc}
\usepackage[margin=1in]{geometry}
\usepackage[parfill]{parskip}

\usepackage[pagewise]{lineno}

\usepackage[british]{babel}
\usepackage[backend=biber,dateabbrev=false,sorting=none]{biblatex}
\renewbibmacro{in:}{}
\usepackage[hidelinks]{hyperref}
\usepackage{csquotes}
\usepackage{graphicx}
\graphicspath{{./images/}}
\usepackage[font=small, labelfont=bf, justification=centering]{caption}
\usepackage{enumitem}
\usepackage{nicefrac}
\usepackage{fancyvrb}
\usepackage{upquote}
\usepackage{booktabs}
\usepackage{multirow}
\usepackage[table,xcdraw]{xcolor}

\title{{\huge Using deep neural networks to improve the precision of fast-sampled particle timing detectors}}

\usepackage{authblk}

\author[1]{Kocot M.}
\author[1]{Misan K.}
\author[1]{Avati V.}
\author[2]{Bossini E.}
\author[1]{Grzanka L.}
\author[3]{Minafra N.}

\affil[1]{AGH University of Krakow, Krakow, Poland}
\affil[2]{INFN Sezione di Pisa, Pisa, Italy}
\affil[3]{Department of Physics and Astronomy, University of Kansas, Lawrence, KS, United States}

\begin{document}

\date{}
\maketitle

The paper has been accepted for publication in Computer Science journal: \url{http://journals.agh.edu.pl/csci}

\section*{Abstract}
Measurements from particle timing detectors are often affected by the time walk effect caused by statistical fluctuations in the charge deposited by passing particles. The constant fraction discriminator (CFD) algorithm is frequently used to mitigate this effect both in test setups and in running experiments, such as the CMS-PPS system at the CERN’s LHC. The CFD is simple and effective but does not leverage all voltage samples in a time series. Its performance could be enhanced with deep neural networks, which are commonly used for time series analysis, including computing the particle arrival time. We evaluated various neural network architectures using data acquired at the test beam facility in the DESY-II synchrotron, where a precise MCP (MicroChannel Plate) detector was installed in addition to PPS diamond timing detectors. MCP measurements were used as a reference to train the networks and compare the results with the standard CFD method. Ultimately, we improved the timing precision by 8\% to 23\%, depending on the detector's readout channel. The best results were obtained using a UNet-based model, which outperformed classical convolutional networks and the multilayer perceptron.

\newpage
\section{Introduction}

Precise time measurements of particles are crucial in many fields, including nuclear medicine and high-energy physics. Designing the most efficient method can be challenging, especially when the required precision is of the order of nanoseconds or picoseconds. Our research focuses on the detectors of the Precision Proton Spectrometer (PPS) \cite{ctpps}, which is a subsystem of the Compact Muon Solenoid (CMS) \cite{cms} detector at CERN's Large Hadron Collider (LHC) \cite{lhc}. At the LHC, protons and ions are accelerated to high energies and are then collided in dedicated interaction points. PPS detects and measures the kinematics of so-called forward protons, which are scattered to small angles after the interaction. Accurate calculation of the particle position and arrival time allows for the precise reconstruction of the particle trajectory and estimation of the interaction position which is crucial for the CMS-PPS subsystem.

\begin{figure}[!tbh]
    \centering
    \includegraphics[width=\textwidth]{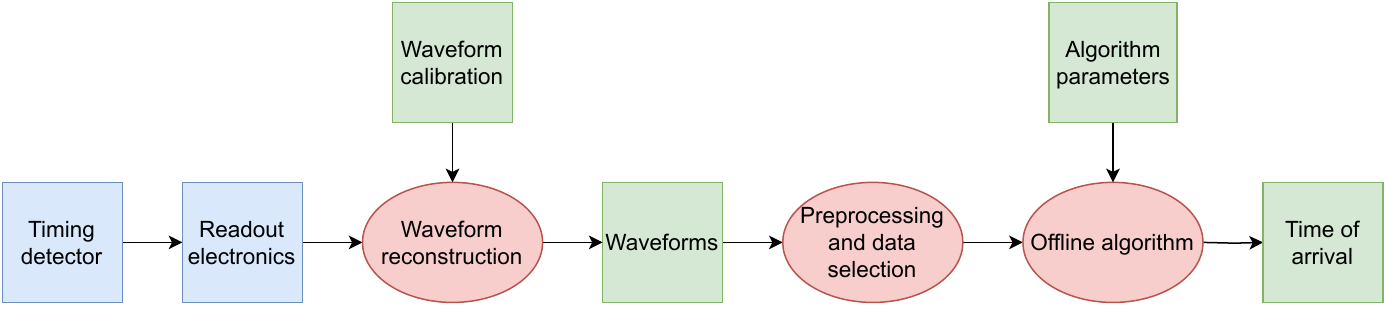}
    \caption{Data flow diagram for the analysis of the timing data (blue: electronics, green: data, red: digital algorithm). Multiple algorithms can be used to retrieve the time of arrival from the waveform data. In this research we focus on digital algorithms working in the offline mode.}
    \label{fig:data_flow}
\end{figure}

The CMS-PPS system uses detectors installed on both sides of CMS, at a distance of approximately 220 m, to perform precise time measurements \cite{cms-timing-run2}. Each detector contains four detection planes with scCVD (single crystal Chemical Vapour Deposition) diamond sensors. When a charged particle passes through a sensor, it generates an electric analogue signal that is later amplified and digitised. In LHC Run 3\footnote{The operating period of the LHC, which started in 2022.}, one of the digitisation techniques uses SAMPIC \cite{sampic}, a fast sampling ASIC (Application-Specific Integrated Circuit), on which we focus in our work. The chip samples and digitises the signal every 156.25\,ps. Proper online or offline\footnote{The word `offline' in this article is used to describe an algorithm working some time after data acquisition, contrary to online algorithms which work in the software or analogue pipeline straight after the data have been acquired.} analysis of these data including a multi-step preprocessing and filtering procedure can be used to compute the particle arrival time with high precision. The data flow during the analysis is depicted in Figure \ref{fig:data_flow}.

\begin{figure}[!tbh]
    \centering
    \includegraphics[width=0.6\textwidth]{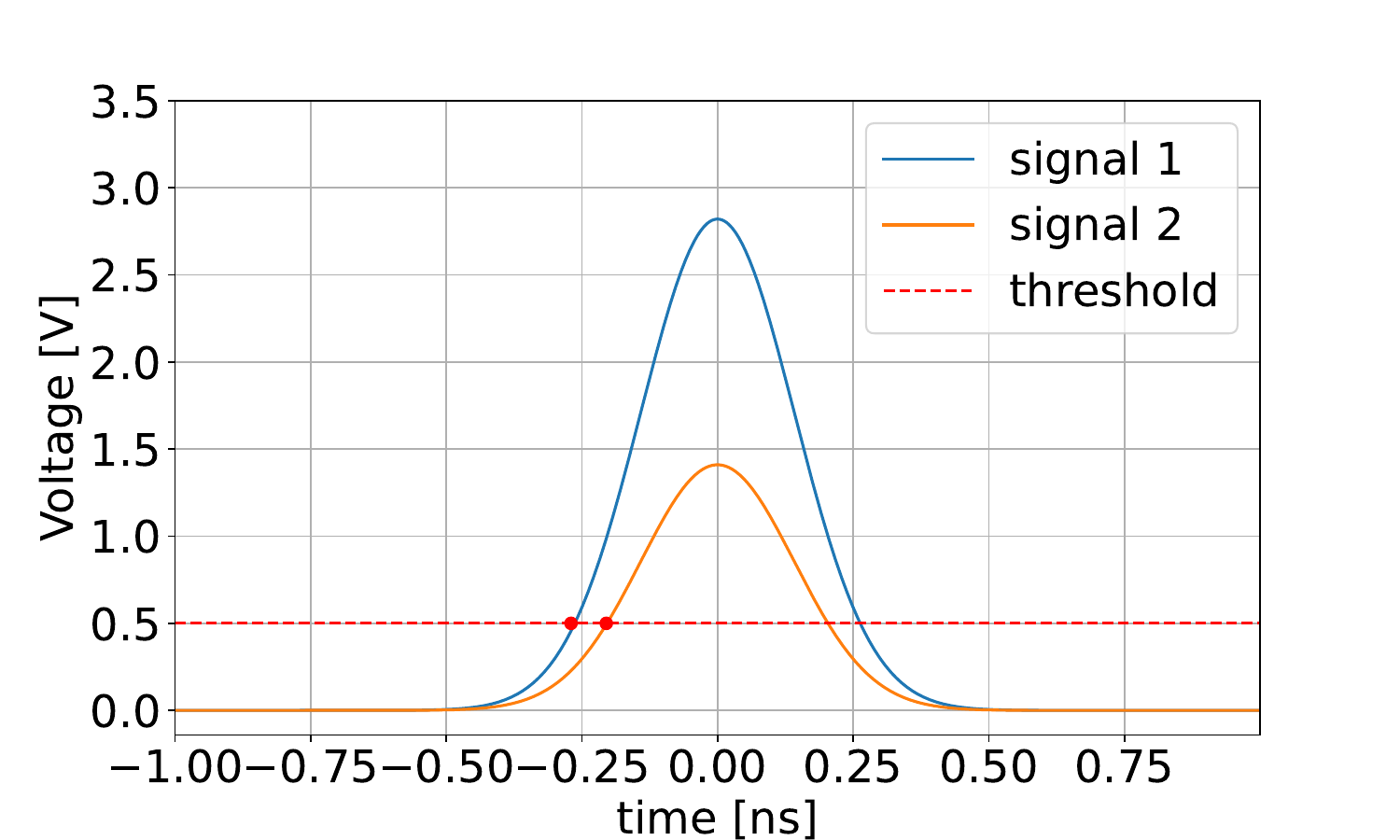}
    \caption{Time walk error illustrated as a difference in the threshold crossing times between two signals with the same shapes but different amplitudes}
    \label{fig:time-walk}
\end{figure}

The accuracy of the arrival time measurement is impacted by two main factors: jitter and the time walk effect. In our work, we focus on minimising the impact of these components on the data from PPS sensors. Jitter is caused by the noise from the signal amplifier. Time walk is the dependence of the measured time on the signal amplitude. It is caused by statistical fluctuations of the charge released in a sensor by a passing particle. This leads to detecting signals with variable amplitudes. Signals with larger amplitudes cross a given threshold earlier than signals with smaller amplitudes. An example of measurements affected by the time walk effect is provided in Figure \ref{fig:time-walk}.

The constant fraction discriminator (CFD) is currently used at CMS-PPS to reduce the impact of the time walk effect and extract the time of arrival timestamps. The CFD is an analytical algorithm and does not use all available samples in a time series. Furthermore, the quality of its results is substantially reduced by the presence of noise and waveform irregularities. To address this issue, we propose a solution that utilises a deep neural network. This network can predict the arrival time of a particle from a sampled time series by using all available samples in a waveform.

\section{State of the art}

Various digital techniques are used to obtain the time of arrival. The classical approach is to use one of the multiple analytical methods. Following the recent trends, machine learning techniques are gaining popularity in this domain, too. This section outlines both of these strategies.

\subsection{Analytical approaches}

The simplest analytical method is the fixed threshold, which extracts the timestamp as the time of crossing a threshold fixed at a specific voltage. The main flaw of the fixed threshold is not taking the time walk effect into consideration at all. 

The most common method used to mitigate the time walk effect is the normalised threshold algorithm, often referred to as the constant fraction discriminator (CFD). This algorithm normalises the waveform amplitude and then applies a fixed threshold. Other techniques include using the signal maximum as the timestamp or extracting only two timestamps and using the time over threshold (TOT) method \cite{timing-methods, timing-methods-2}.

Due to its simplicity and relatively high performance, the Constant Fraction Discriminator (CFD) is considered the most reliable choice \cite{edo-thesis}. It is used in both test setups and running experiments, such as the CMS-PPS system at the LHC. Originally, the CFD was devised as an analogue device. However, we use it as an offline algorithmic solution to measure the arrival time of a particle given a very fast electrical pulse. By mitigating the error introduced by the time walk effect, the CFD allows for very accurate timing measurements. Given the excellent properties of the CFD and its common usage in the field, we selected this algorithm as the baseline for our numerical experiments.

\subsection{Machine learning methods}

Machine learning techniques are widely used in high energy physics. Common use cases include monitoring data quality by identifying outliers \cite{dqm-ml} and particle track reconstruction \cite{track-reco}. The short execution time of machine learning methods makes them useful for the high level trigger reconstruction task \cite{hlt-ml}.

Although some supervised machine learning techniques, specifically deep neural networks, show promising results in time series analysis and timestamp prediction \cite{time-series-1}, they are seldom used to predict the time of arrival and have never been utilised for this purpose in the CMS-PPS subsystem.

The most extensive tests of neural networks in the domain of computing the time of arrival have been performed for MRPC (Multigap Resistive Plate Chamber) detectors. The research showed that multilayer perceptrons, LSTM (Long Short-Term Memory) recurrent neural networks or their combinations can be successfully used with the signals from a particle timing detector \cite{mrpc-1, mrpc-2}. The interest is high in medical applications, too. Various convolutional architectures, mainly UNet-based, are used to calculate the time of flight in PET detectors \cite{pet} and to tag ECG diagrams \cite{ecg-1, ecg-2}. While these problems are different from the one discussed in this paper, they still require time series tagging, which is at the core of our problem.

\section{Dataset}

This section describes the data source and preprocessing steps required to construct the dataset used in this work. We also provide a detailed description of the version of the CFD algorithm, which is used during the data preprocessing procedure.

\subsection{Data source}

We constructed a dataset using the data acquired at the test beam facility in the DESY-II synchrotron in 2020 \cite{desy}. The facility hosted the PPS diamond timing detectors, as well as an MCP-PMT (Microchannel Plate Photomultiplier Tube) detector. The sensors were connected to the SAMPIC readout chip. The voltage time series sampled by SAMPIC had a fixed length of 64 samples within a 10 ns time window. Typically, the time window was long enough to capture the entire MCP signal. However, with diamond sensors, the signal is typically longer, with a wider trailing edge compared to the leading edge. As a result, in most cases, 10\,ns was enough to fully capture only the leading edge of a signal.

\begin{figure}[!tbh]
    \hfill
    \begin{minipage}[b]{0.45\textwidth}
        \includegraphics[width=\textwidth]{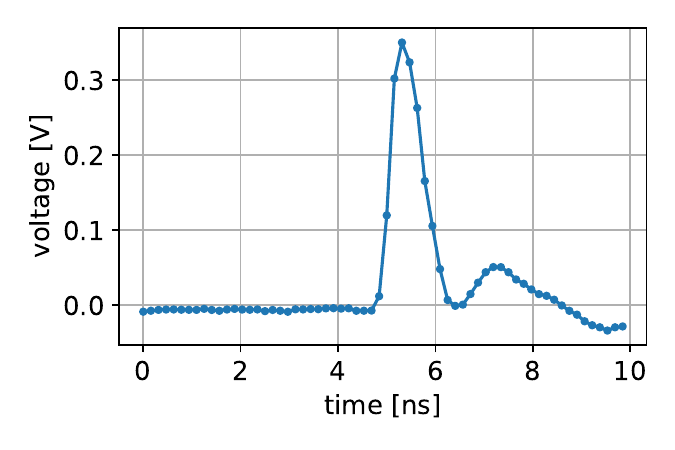}
    \end{minipage}
    \hfill
    \begin{minipage}[b]{0.45\textwidth}
        \includegraphics[width=\textwidth]{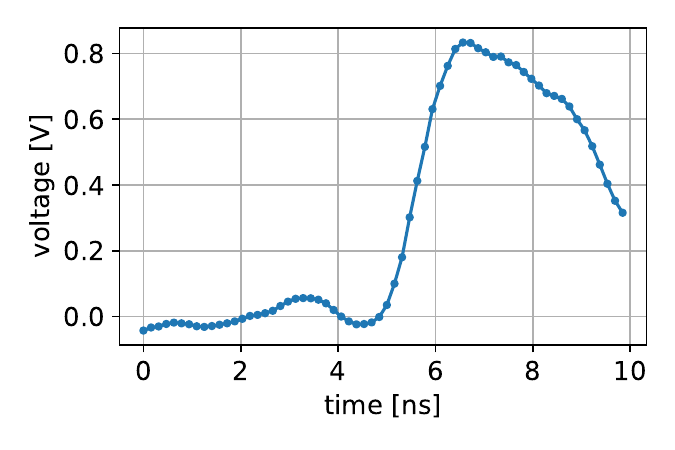}
    \end{minipage}
    \hfill \hfill
    \caption{Example waveforms from the MCP (left) and a diamond detector (right).}
    \label{fig:example_waveforms}
\end{figure}

The expected precision of the PPS diamond sensors was 50-100\,ps, while the MCP timing precision was measured to be around 10\,ps \cite{desy}. Considering its performance, MCP readouts were a perfect source of ground-truth information for our experiments. We present examples of waveforms acquired using the MCP and a diamond sensor in Figure \ref{fig:example_waveforms}.

\subsection{Preprocessing}

Multiple preprocessing steps were required for the data acquired from the DESY beam. Firstly, the samples were inverted to compensate for negative signals. This operation resulted in a rising edge in a signal indicating a particle.

\begin{figure}[!tbh]
    \hfill
    \begin{minipage}[b]{0.45\textwidth}
        \includegraphics[width=\textwidth]{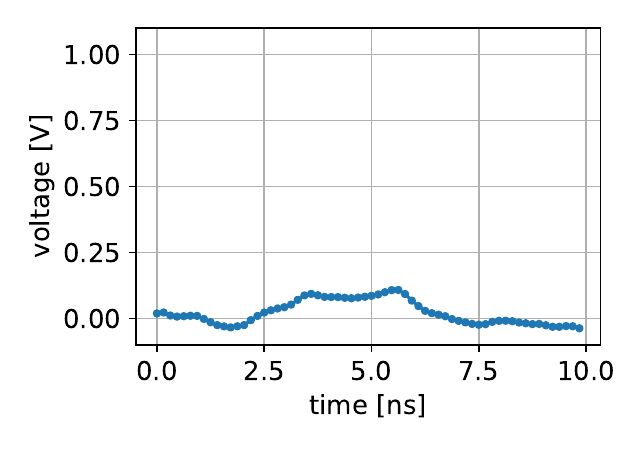}
    \end{minipage}
    \hfill
    \begin{minipage}[b]{0.45\textwidth}
        \includegraphics[width=\textwidth]{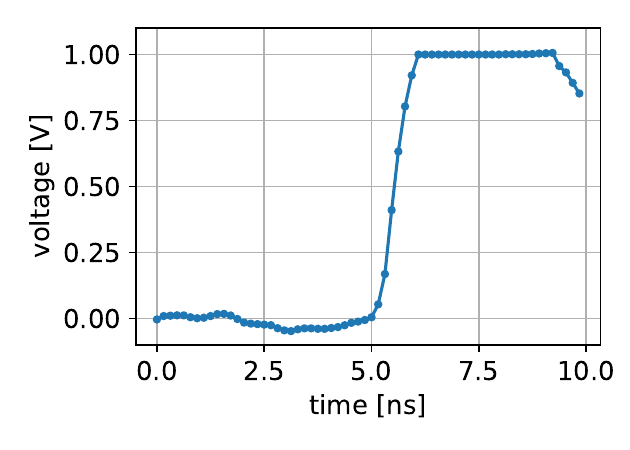}
    \end{minipage}
    \hfill \hfill
    \caption{Examples of noisy (left) and saturated (right) events acquired using the diamond sensor}
    \label{fig:noisy_saturated_examples}
\end{figure}

The next preprocessing step was to eliminate noisy and so-called saturated events. Noisy signals are too weak to analyse, while saturated events occur when the voltage exceeds the amplifier's dynamic range, resulting in the signal being capped at a certain level. Examples of such events are shown in Figure \ref{fig:noisy_saturated_examples}. This means that the true amplitude cannot be easily retrieved from the signal, and the CFD cannot produce accurate reference values. We excluded noisy and saturated events from the analysis to focus solely on the comparison between deep learning models and the CFD. 

\begin{figure}[!tbh]
    \hfill
    \begin{minipage}[b]{0.325\textwidth}
        \includegraphics[width=\textwidth]{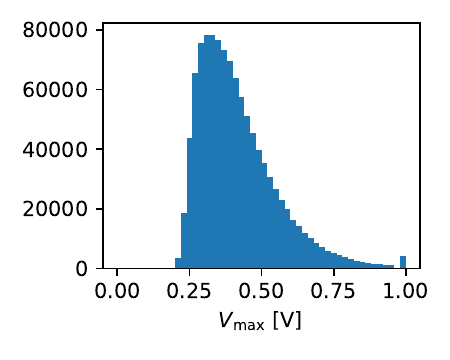}
    \end{minipage}
    \hfill
    \begin{minipage}[b]{0.325\textwidth}
        \includegraphics[width=\textwidth]{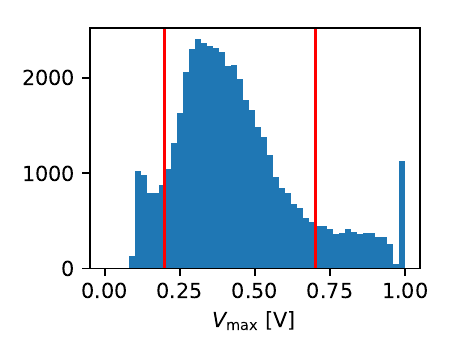}
    \end{minipage}
    \hfill
    \begin{minipage}[b]{0.325\textwidth}
        \includegraphics[width=\textwidth]{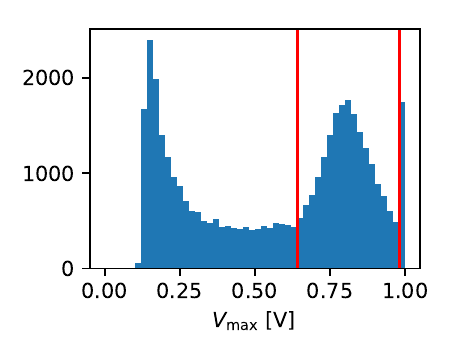}
    \end{minipage}
    \hfill \hfill
    \caption{Maximum voltage histograms for MCP (left) and two selected diamond detectors (middle and right). The red lines on the histograms of the diamond detectors indicate the minimum and maximum amplitude cuts. It is important to note that the histograms look different for each detector, and therefore, it was necessary to find the amplitude cut values separately for each of them.}
    \label{fig:vmax_histograms}
\end{figure}

The data filtering was done mainly using amplitude histograms. Amplitude (i.e. maximum voltage) was plotted on a histogram for each event. Typically, noisy events appear on the left side of the histogram, while saturated events appear on the right side. Therefore, it is simple to filter out such events by selecting minimum and maximum amplitudes and discarding any events that fall outside this range. The minimum and maximum cuts were determined manually by analysing the histogram shapes. Figure \ref{fig:vmax_histograms} shows the maximum voltage histograms for the MCP and two selected diamond detectors. Due to its excellent waveform quality and low ratio of saturated events (visible as a small peak on the right side of its histogram), MCP did not require filtering.

In the original dataset, the distribution of signal rising edge timestamps was centred around a single value in the middle of the time window. Using such a dataset would make it highly likely for the networks to overfit and collapse the predictions to the same timestamp for any input data. To avoid this, the signals were trimmed from 64 to 48 samples by removing 16 samples from the edges of the window. Specifically, up to 16 first samples were cut from each signal, and only the next 48 samples were kept. The number of first samples to drop was chosen randomly to smear the timestamp distribution.

Another issue was the fact that the waveforms had varying baselines and amplitudes. They had to be normalised in order to be properly processed by the neural networks. At first, the baseline was calculated as the mean value of the first 20 samples. Then, it was subtracted from the voltage values in the time series. Afterwards, the waveforms were divided by their maximums to normalise the amplitudes. The same steps are used in our version of the CFD algorithm and are visualised in Figure \ref{fig:cfd} (left).

\subsection{Final dataset}

\begin{figure}[!tbh]
    \hfill
    \begin{minipage}[b]{0.45\textwidth}
        \includegraphics[width=\textwidth]{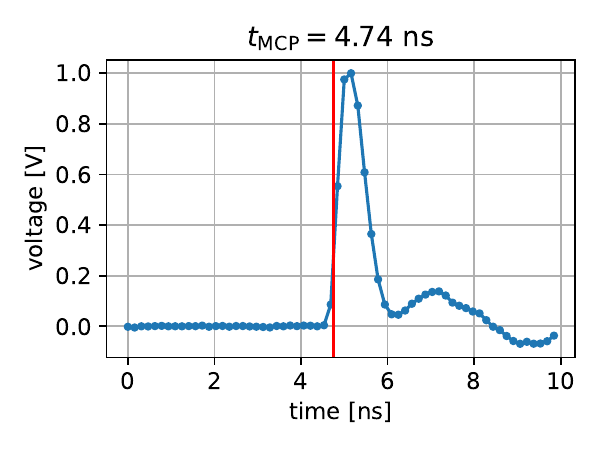}
    \end{minipage}
    \hfill
    \begin{minipage}[b]{0.45\textwidth}
        \includegraphics[width=\textwidth]{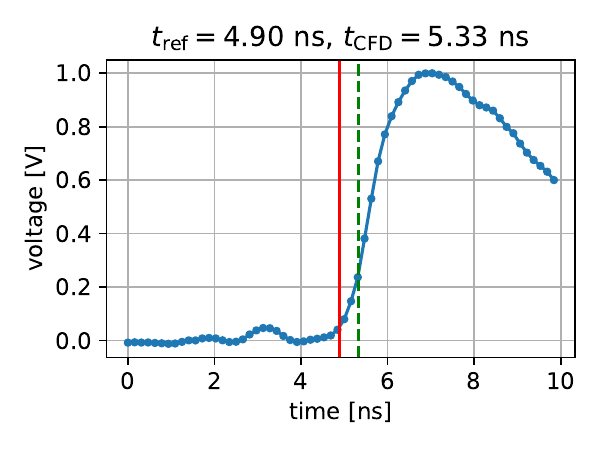}
    \end{minipage}
    \hfill \hfill
    \caption{Example from the final dataset. Left: waveform from MCP. $t_\mathrm{MCP}$ (red line) is computed using the CFD. Right: waveform from a diamond detector. $t_\mathrm{ref}$ (red line) is the neural network's reference time computed using the CFD for the MCP waveform. The green dashed line represents the time computed using the CFD with the waveform from the diamond detector ($t_\mathrm{CFD}$). It is not included in the dataset and is shown only for visualisation purposes.}
    \label{fig:dataset_example}
\end{figure}

In order to train the neural networks, we needed both time series and ground-truth (reference) time. Therefore, we constructed a dataset that only included events with corresponding readouts from both the MCP and a diamond sensor. The ground-truth timestamps were obtained from the MCP signals using the CFD method explained below. Due to the high quality of the MCP measurements, this approach was sufficient and provided the necessary timing precision for the training process. Figure \ref{fig:dataset_example} presents a normalised waveform from the final dataset, along with a corresponding MCP signal and the reference timestamp. The slight difference between the reference time shown on the MCP and diamond detector waveforms is due to the relative difference in the times of the first samples of both signals. This difference must be taken into account in the calculations to avoid bias.

The final dataset consisted of approximately 500,000 waveform entries and their corresponding reference timestamps. After removing irrelevant information, performing data filtering and preprocessing, the size of the dataset was reduced from around 5.5 GB to 126 MB.

\subsection{Constant Fraction Discriminator}

The CFD algorithm used in this research is a modified version of the normalised threshold algorithm.

\begin{figure}[!tbh]
    \hfill
    \begin{minipage}[b]{0.45\textwidth}
        \includegraphics[width=\textwidth]{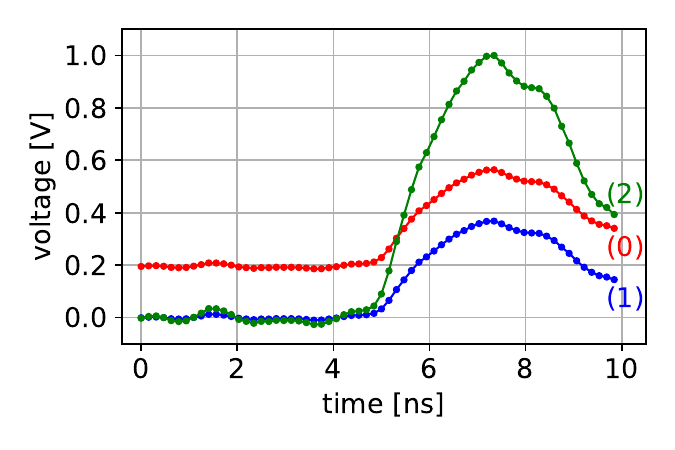}
    \end{minipage}
    \hfill
    \begin{minipage}[b]{0.45\textwidth}
        \includegraphics[width=\textwidth]{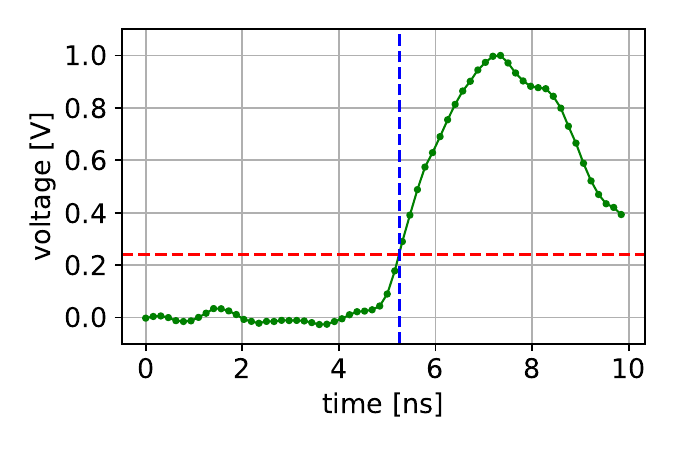}
    \end{minipage}
    \hfill \hfill
    \caption{Depiction of the CFD algorithm. Left: (0) before normalisation, (1) baseline subtraction, (2) division by maximum; right: when the normalisation is done, the timestamp can be found using the fixed threshold algorithm.}
    \label{fig:cfd}
\end{figure}

First, we normalise a time series using the same strategy as during data preprocessing, which involves baseline subtraction and division by the amplitude. Next, we calculate the time of arrival as the moment when the series crosses a chosen voltage threshold. To determine the exact timestamp, we apply a linear interpolation between the point before and the point after the threshold crossing. The chosen threshold is a fraction of the normalised amplitude, which ensures that the crossing point's dependence on the amplitude is removed. Figure \ref{fig:cfd} illustrates these steps.

\section{Network architectures}

Our goal was to improve the timing precision using deep neural networks. We aimed to demonstrate that our methods achieve better results in terms of particle arrival time precision than the CFD. We started from a multilayer perceptron (MLP) and progressively increased the overall complexity of the network structure by using regular convolutional architectures and UNet-based \cite{unet} networks. We ran a hyperparameter tuning algorithm for each network type and selected the best candidates. Below, we briefly describe the training configuration, tuning method and hyperparameter options. We also depict the best-performing models.

\subsection{Training configuration}

We selected a single detector readout channel, i.e. a single diamond detector, for our primary tests. After preprocessing, we obtained 15,675 and 3,919 entries in the training and test sets, respectively. However, at this point, we left the test set for the final performance assessment. The models were trained using the Adam \cite{adam} optimiser with an adaptable learning rate which was reduced on learning curve plateaus. 

As the output of the MCP and regular convolutional models was just a single number (the predicted timestamp), their metric was the squared error between the predicted and ground-truth values. In the case of the UNet-based model, it was trained to output a heatmap, so its error was calculated as the mean squared error between the predicted heatmap and the ground-truth one. The ground-truth heatmap was generated as a Gaussian with the mean at the true timestamp and a small standard deviation of one sampling step, i.e. 156.25\,ps, following \cite{tompson2014joint}. The final UNet timestamp could be retrieved as the mean of a Gaussian fitted to the output vector. The training process stopped when no improvements in the loss function were observed, following the early stopping method. This aimed to minimise the impact of overfitting \cite{early-stopping-overfitting}.

\subsection{Hyperparameter tuning procedure}

Our hyperparameter tuning procedure consisted of two steps. First, we used a tuner algorithm to select the most promising models. Then, we performed cross-validation to determine the best one. The entire procedure utilised only the training set, with the same train-validation splits for every network type. We reserved the test set for final performance estimation.

For the first step, we chose one of the most common hyperparameter tuners, KerasTuner \cite{keras-tuner}. It is capable of selecting the top-N best models given an optimisation algorithm. We chose the Bayesian optimiser, which is an improved version of the grid search and random search algorithms. It estimates the loss function versus the hyperparameter values, and samples the hyperparameter sets according to that distribution. For each network type, we ran 40 iterations of KerasTuner, testing 40 different hyperparameter sets. Each set was trained using 80\% of the original training set, while the remaining 20\% was used for validation. To improve the quality of the results and filter out unstable models, we used two executions per trial, meaning that the result of a model was calculated as an average of the loss values from these two, separate executions of training and validation.

In the second phase, we used the top 5 models outputted by the tuner and performed 5-fold cross-validation using only the training set. The folds were consistent across all network types, and each fold value was an average of three trials. The final model for a given network type was chosen based on the mean and standard deviation of the cross-validation results.

The hyperparameter tuning procedure was run on the High-Performance Computing GPU cluster. The computations took from one to four hours, depending on the network architecture.

\subsection{Multilayer perceptron}

The main goal of tuning the multilayer perceptron (MLP) architecture was to find the optimal number of hidden layers and neurons. To avoid using a separate hyperparameter for the number of neurons in each layer, we assumed that either the same number was used or that the number was divided by 1.5 or 2 for every consecutive dense layer. The final layer always had only one neuron, since it was the output of the model and represented the predicted timestamp.

In addition to these parameters, we also tested the effects of adding batch normalisation \cite{batch-normalisation} and dropout \cite{dropout}. Batch normalisation could be added either after every dense layer or not at all, and could also be added independently after the input layer. Finally, dropout was set to either 0, 0.2, or 0.5. An activation function was applied after every dense layer, except for the last one, and was fixed to ReLU \cite{relu}.

\begin{figure}[!tbh]
    \centering
    \includegraphics[width=0.5\textwidth]{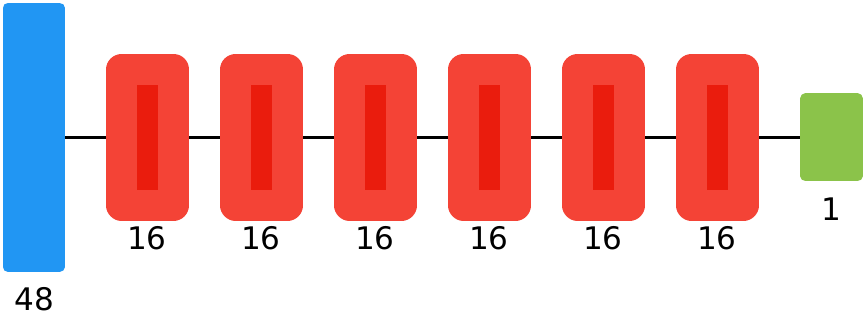}
    \includegraphics[width=0.5\textwidth]{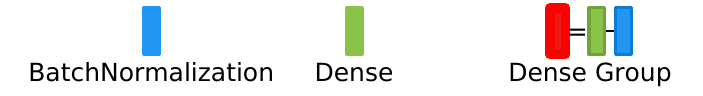}
    \caption{Optimal MLP model \cite{net2vis}}
    \label{fig:mlp}
\end{figure}

The best models returned by the tuning procedure almost always included batch normalisation after every dense layer and the input but did not use dropout. There were no clear patterns for the rest of the hyperparameters. The final MLP architecture is shown in Figure \ref{fig:mlp}.


\subsection{Convolutional Network}

The convolutional neural network (CNN) consisted mainly of convolutional layers, which are commonly used in image and time series processing tasks. Unlike dense layers, convolutional layers can learn the relations between neighbours, such as time series samples located next to each other.

Typically, the number of filters increases with each consecutive convolutional layer. In our case, it was multiplied by 2. However, up to three convolutional layers could be used in sequence before increasing the filter count. We refer to this group as a convolutional block. The number of blocks was another hyperparameter, ranging from 1 to 4. Only the number of filters in the first block was optimised, as the numbers in the following blocks could be inferred from the number in the first block. All convolutional layers had kernels of small size: 3. A single dense layer was placed at the end of the network so that the network could output a single number. A small MLP could be inserted between the convolutional part of the network and the final dense layer, parametrised similarly to the full MLP architecture. However, its depth was limited to 3.

In addition to the core layers, batch normalisation was parametrised similarly to the MLP. Dropout could be applied after each dense layer in the MLP part. For convolutional layers, we used spatial dropout \cite{spatial-dropout} instead of regular dropout. Its rate could be set to 0.0, 0.1, or 0.2.

\begin{figure}[!tbh]
    \centering
    \includegraphics[width=0.75\textwidth]{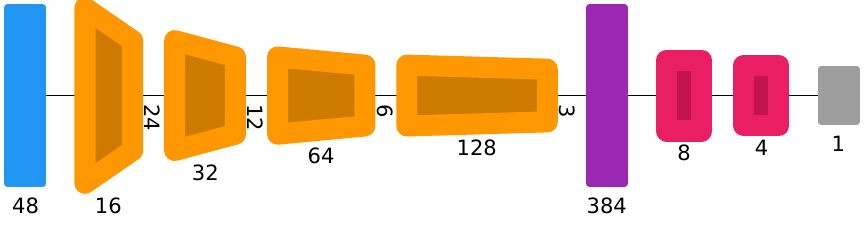}
    \includegraphics[width=1.0\textwidth]{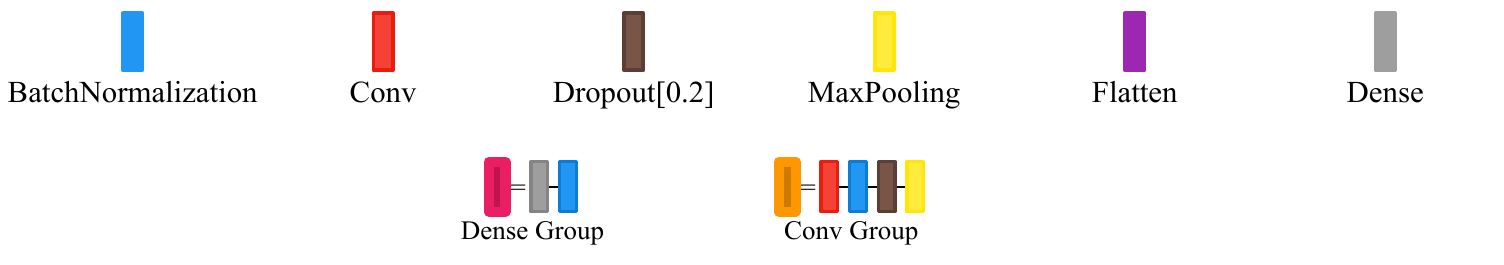}
    \caption{Optimal CNN model \cite{net2vis}}
    \label{fig:convnet}
\end{figure}

Batch normalisation was used in all of the models returned by the tuner, while the MLP dropout was always set to 0.0. No visible patterns were observed for other hyperparameters. The final convolutional architecture is shown in Figure \ref{fig:convnet}.

\subsection{UNet} 

The last architecture we used, UNet, is characterised by the U shape of its architecture. It is composed of an encoder and a decoder. The encoder extracts relevant features from the input, while the decoder uses those features to build a vector of the input shape and highlight relevant spots, such as predicted timestamps in time series processing. The UNet architecture takes advantage of skip connections to amplify the importance of initial features in the decoder. Thanks to the segmentation and noise reduction capabilities of UNet, we expected it to be a good candidate for our task.

\begin{figure}[!tbh]
    \centering
    \includegraphics[width=1.0\textwidth]{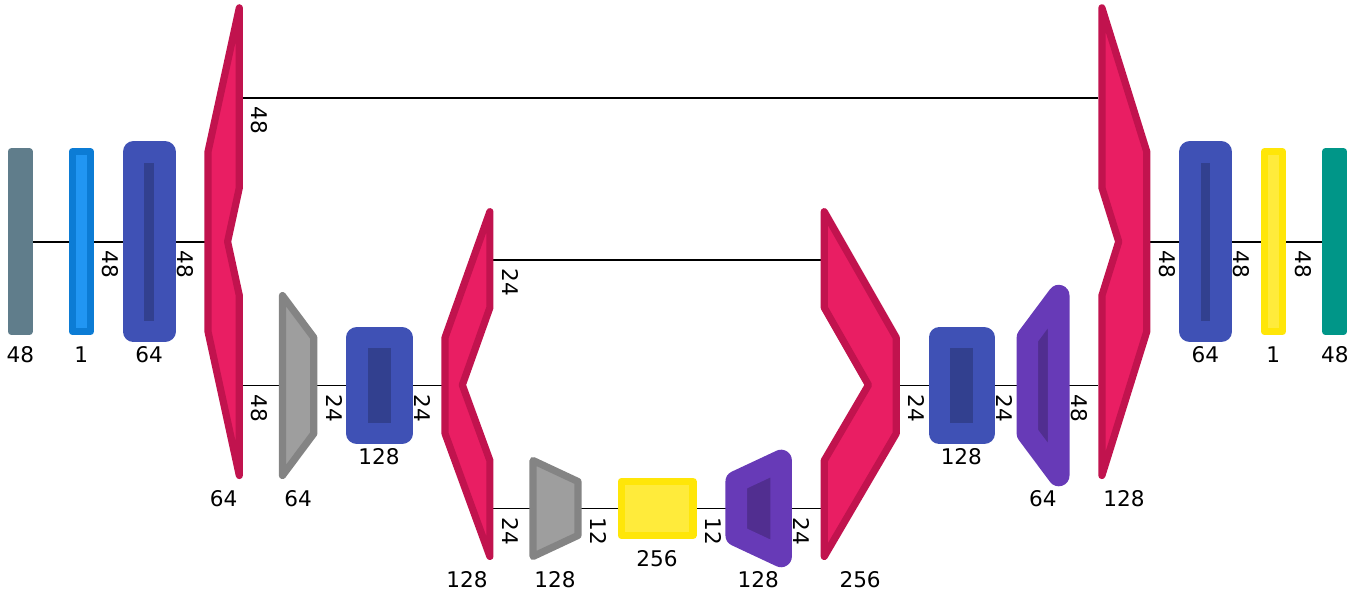} \\
    \includegraphics[width=0.87\textwidth]{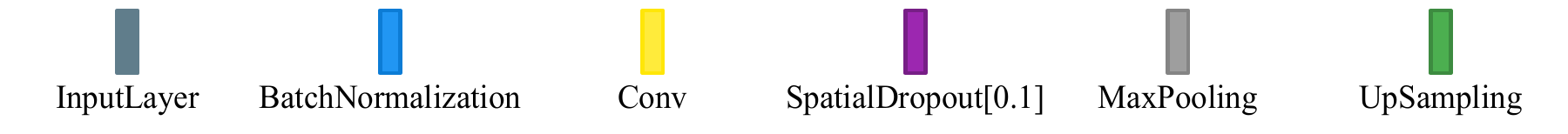} \\
    \includegraphics[width=1.0\textwidth]{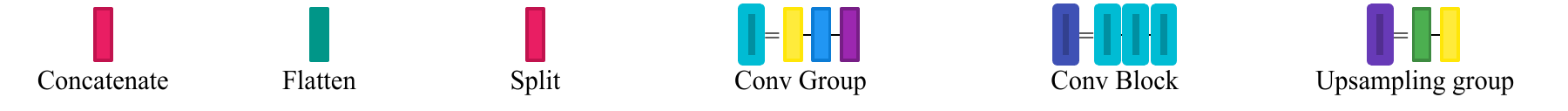}
    \caption{Optimal UNet model \cite{net2vis}}
    \label{fig:unet}
\end{figure}

The main hyperparameter for our model of the network depth, measured in UNet blocks. The encoder and decoder were symmetrical and contained the same number of blocks. An encoder block consisted of one to three convolutional layers followed by a max pooling layer. A decoder block started with deconvolution, which we implemented through upsampling and a convolutional layer with a kernel size of one (all other convolutional layers had a kernel size of 3). The output of a decoder block was concatenated with the output from the corresponding block in the encoder through a skip connection. Finally, one to three convolutional layers were used. As before, batch normalisation and spatial dropout could be added after every convolutional layer with a kernel size of 3. Batch normalisation could also be used after input. The final UNet architecture is depicted in Figure \ref{fig:unet}.

\section{Results}

This section begins with a description of the method used to assess the timing precision of either the CFD or a deep neural network. Then, we compare the results obtained with the CFD to those obtained with the best neural network using data from a single readout channel of the detector. We also provide a description of other performance tests we performed, including the tests on other readout channels.

\subsection{Precision assessment method}

To measure the time precision of a detector, a typical method is to compare its measurements with those of a `reference' detector that has much better time precision. The reference detector is placed on the same beam line to detect the same particles. The mean of the differences between the tested detector and the reference detector represents a constant offset. Although neural networks can learn to have a mean close to zero, the CFD mean is expected to be shifted due to the method's inability to adjust to inconsistent signal characteristics. The precision of the time measurement is represented by the standard deviation of the differences.

In our case, MCP served the role of the reference detector. What is more, to reduce tail effects, we fitted a Gaussian curve to the histogram of the time differences and used the standard deviation of the Gaussian as the precision measure (as shown in Figure \ref{fig:diff}).

\subsection{The optimal architecture}

We performed cross-validation on the best models, one from each network type. Instead of computing loss values, we used the evaluation method described above to compute the results. The computations were performed only on the training portion of the dataset. The results are shown in Table \ref{tab:optimal-architecture}. As expected, UNet had the smallest (the best) average precision value. It was also the most complex model with the biggest number of parameters. Interestingly, the model had more parameters than the number of training samples. This is typical for neural networks as they often use more parameters than the minimum required number. While this can make the network susceptible to overfitting, with proper training, overfitting can be avoided. The early stopping method we employed is the best approach to address this issue. Additionally, spatial dropout was applied to improve performance at the expense of further increasing the number of parameters. Surprisingly, the MLP model was the most stable, with the smallest standard deviation of results for each fold.

\begin{table}[!tbh]
\centering
\begin{tabular}{llll}
\toprule
Architecture & Mean {[}ps{]} & Std {[}ps{]} & Parameter count \\ \midrule
\rowcolor[HTML]{EFEFEF} 
MLP          & 63.90         & 0.85         & 2737            \\ 
CNN          & 62.83         & 1.34         & 36865           \\ 
\rowcolor[HTML]{EFEFEF} 
UNet         & 60.71         & 1.19         & 456965          \\ \bottomrule
\end{tabular}
\caption{Comparison of the precisions achieved by the optimal models in the cross-validation procedure. In addition to the cross-validation scores, the number of parameters used by each network is reported, too.}
\label{tab:optimal-architecture}
\end{table}

\begin{figure}[!tb]
    \hfill
    \begin{minipage}[b]{0.45\textwidth}
        \includegraphics[width=\textwidth]{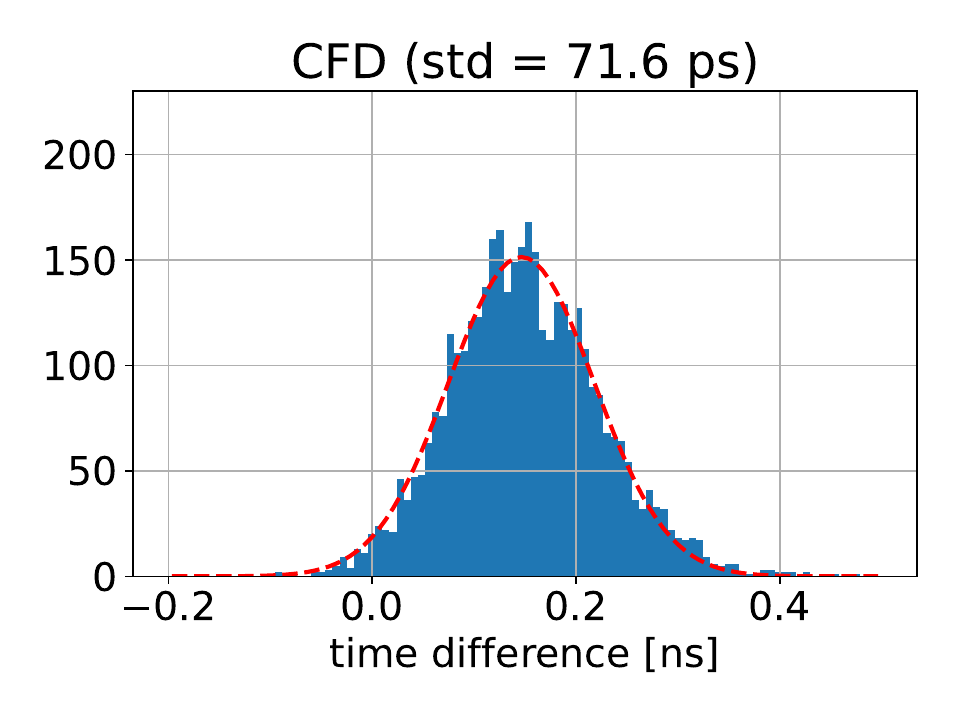}
    \end{minipage}
    \hfill
    \begin{minipage}[b]{0.45\textwidth}
        \includegraphics[width=\textwidth]{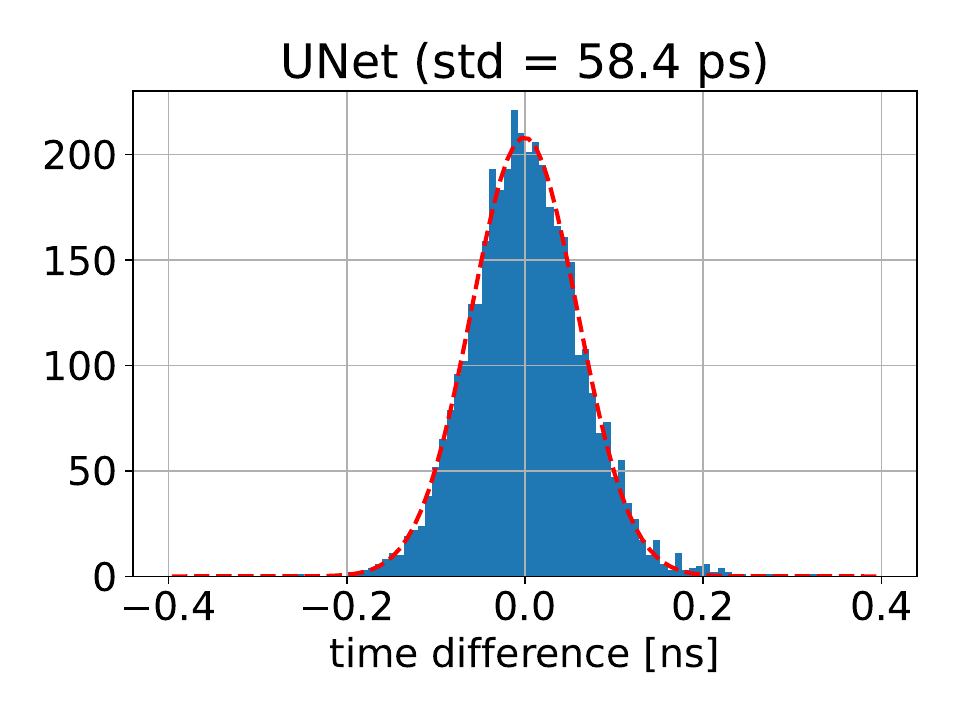}
    \end{minipage}
    \hfill \hfill
    \caption{Difference histograms for the CFD and our best-performing, UNet-based model}
    \label{fig:diff}
\end{figure}

To evaluate the final performance of our solution, we used the test set which was composed of data from the same readout channel as the training set. Figure \ref{fig:diff} shows the difference histograms for the CFD and our best-performing neural network. The network's histogram is visibly narrower, indicating better precision. 

\subsection{Adjusting the data to the LHC conditions}

Due to limited available bandwidth in the PPS setup at the LHC, the SAMPIC time series consist of only 24 samples. To validate the networks under these conditions, we trimmed the original time series from 64 to 24 samples. First, we smeared the timestamp distribution by randomly removing up to 10 samples from the beginning of the series, retaining only the next 56 samples. We then selected the 24 samples from the middle to ensure that the most important part of the signal was preserved. We made slight adjustments to the network architectures to accommodate the smaller input size (24 instead of 48). We performed the same tests as before and obtained similar results. We were able to improve the precision from 73.3\,ps to 62.1\,ps standing for 15\%, which is promising for using deep learning with LHC data.

\subsection{Tests on many channels}

In the previous sections, we only used data from a single detector channel. However, we also tested data from other channels. We first trained the optimal network on each channel separately and then tested it on the same channel which was used for training. We achieved precision improvements ranging from 8\% to 23\% compared to the CFD. We also investigated if the network could be trained on one channel and tested on another, or even trained on all available data while maintaining the train-test split. We present the results for selected, representative channels in Table \ref{tab:many-channel-precisions}.

\begin{table}[!tb]
\centering
\begin{tabular}{@{}rrrrrrrr@{}}
\toprule
\multicolumn{1}{c}{} & \multicolumn{6}{c}{test channel} \\ \cmidrule(l){2-8} 
\multicolumn{1}{c}{\multirow{-2}{*}{training channel}} & \multicolumn{1}{c}{10} & \multicolumn{1}{c}{16} & \multicolumn{1}{c}{17} & \multicolumn{1}{c}{22} & \multicolumn{1}{c}{25} & \multicolumn{1}{c}{27} & \multicolumn{1}{c}{31} \\ \midrule
\rowcolor[HTML]{EFEFEF} 
10  & \textbf{13\%} & 10\%          & 13\%          & 7\%           & -1\%          & -23\%         & -7\%          \\
16  & 6\%           & \textbf{23\%} & 16\%          & 9\%           & -22\%         & -9\%          & 3\%           \\
\rowcolor[HTML]{EFEFEF} 
17  & 7\%           & 17\%          & \textbf{19\%} & 9\%           & -3\%          & 8\%           & -6\%          \\
22  & 4\%           & 14\%          & -4\%          & 11\%          & -84\%         & -51\%         & 7\%           \\
\rowcolor[HTML]{EFEFEF} 
25  & 4\%           & 4\%           & 7\%           & 4\%           & \textbf{12\%} & 8\%           & -4\%          \\
27  & -13\%         & -10\%         & 4\%           & -16\%         & 4\%           & \textbf{19\%} & -17\%         \\
\rowcolor[HTML]{EFEFEF} 
31  &  2\%          & 13\%          & 10\%          & 7\%           & -7\%          & -7            & \textbf{16\%} \\ \midrule
all & 8\%           & 22\%          & 14\%          & \textbf{12\%} & 9\%           & 17\%          & 14\%          \\ \bottomrule
\end{tabular}
\caption{Precision improvements with respect to the CFD obtained with many detector channels involved. We selected seven, representative channels and highlighted the best precision for each one.}
\label{tab:many-channel-precisions}
\end{table}

\begin{figure}[!tb]
    \centering
    \includegraphics[width=0.9\textwidth]{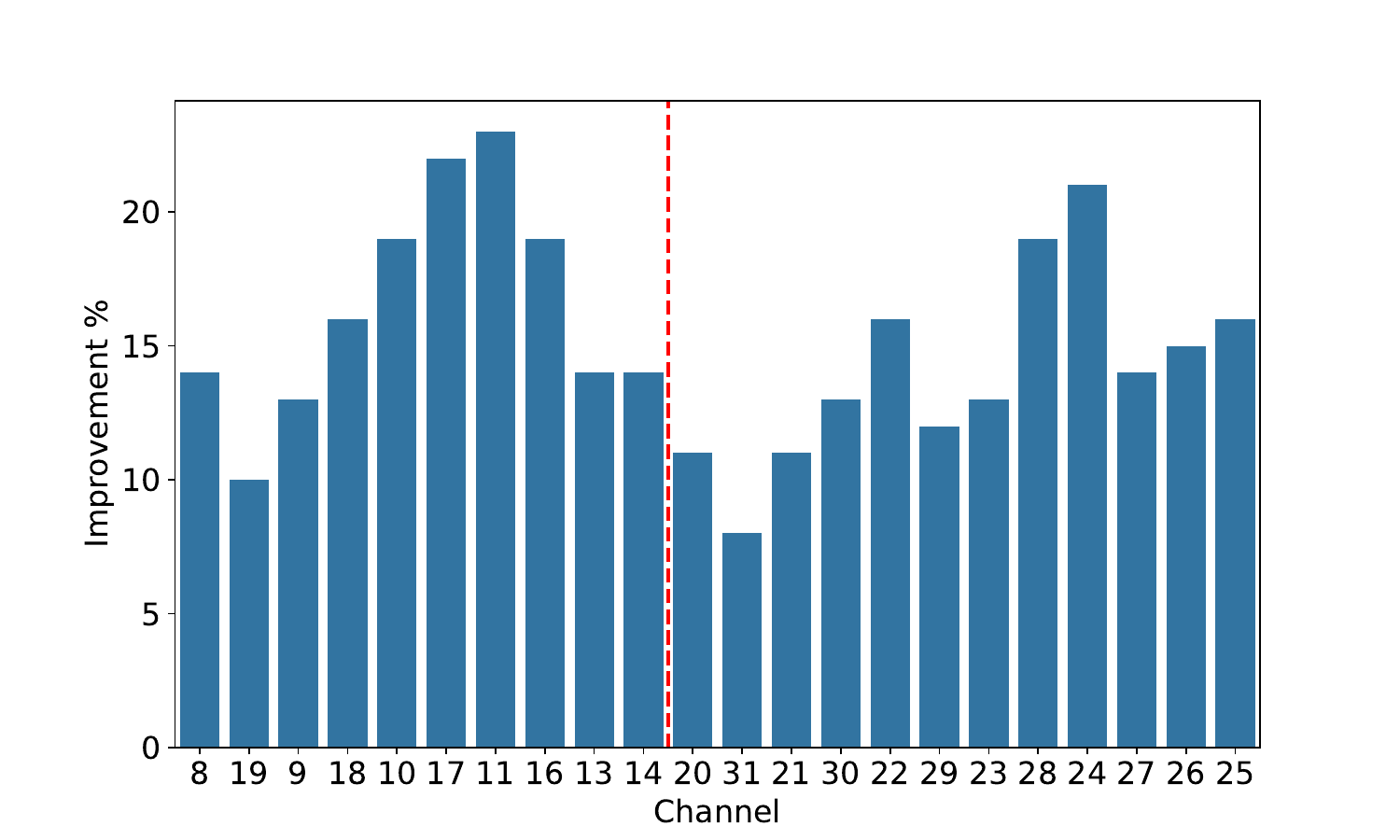}
    \caption{Improvements with respect to the CFD for the channels we explored. The channel order follows the physical set-up of the test beam experiment. The red dashed line divides the channels from two separate planes of diamond sensors used in the experiment.}
    \label{fig:many-channel-improvements}
\end{figure}

These results show that, typically, in order to achieve the highest precision for a given channel, the network needs to be trained using data from that channel specifically. If trained on one channel and tested on another, the network might perform worse than the CFD, resulting in negative entries in Table \ref{tab:many-channel-precisions}. This shows that even though the same sensor is used, the collected data differ significantly between channels. Channel 22 is the only channel for which the network trained on all channels was able to achieve slightly better precision than the network trained on that specific channel. This may be due to an unfavourable train-test split for the channel 22 data. We present the improvements for all the channels we explored using the networks trained on the particular channels in Figure \ref{fig:many-channel-improvements}. The data were available for channels from 8 to 31. We do not report the improvements for channels 12 and 15, as the waveforms were too noisy.

It is worth noting that we did not test the network on data from all available channels at once. This would be difficult due to the different mean values of the difference histograms in various channels. Instead of a single Gaussian, we would have a group of smeared Gaussians, which would make it impossible to retrieve the true detector precision. Therefore, even when the network was trained using data from all channels, we tested it separately for each channel.

\section{Conclusion}

We demonstrated that deep neural networks can be used to compute the time of arrival of particles taking samples voltage signals at input. In fact, these networks can improve timing precision compared to the most commonly used algorithm, the CFD. In the base numerical experiment, we were able to improve precision by 17\%. It is a significant value considering that we did not make any modifications the detector setup, but just used a different algorithm for computing the time of arrival. Other readout channels also showed improvements ranging from 8\% to 23\%. We found that networks based on the UNet architecture yield the best results among the models we investigated. However, we did not test recurrent networks, which are also expected to perform well in this kind of problem. We leave that for further research. Nonetheless, even the simplest network architecture we tested -- MLP -- enabled us to calculate the arrival time with noticeably better precision than the CFD.

Neural networks have a wide range of applications in high energy physics. For example, some types of neural networks are used to evaluate the quality of waveform-like experimental data, thereby improving the detection of outliers and bad data \cite{dqm-ml}. Additionally, the resolution of pattern recognition algorithms is improved for detectors with complex geometries \cite{track-reco}. This work contributes to the evaluation of neural networks in high-energy physics. The method proposed in this manuscript broadens their applicability and increases the precision of timing detectors. Deep neural networks have been tested in similar applications. It has been shown that they can be used in the prediction of arrival time (\cite{mrpc-1, mrpc-2}), or more generally in the annotation of time series (\cite{ecg-1}). This proves that our findings are not just a coincidence. However, it is worth noting that neural networks can achieve high accuracy only on data similar to the training dataset.

The method described in \cite{mrpc-2} differs from our approach in terms of the detector architecture and the usage of simulated data in network training. In contrast, our work relies entirely on experimental data and a more precise source of reference data in the form of an MCP-PMT detector.

It is important to note that our reference data were not flawless. The MCP precision was assessed to be 10 ps, which is much better than that of diamond sensors (about 50-100 ps). Nevertheless, MCP signals are not perfect and introduce a small degree of uncertainty. Using the CFD further worsens the reference precision. The resulting error is random and can cause some events to contradict each other during neural network training. For instance, reference timestamps may vary for waveforms that look identical. As a result, the final precision of the neural network was negatively impacted.

In addition to the time of arrival study, the procedure developed in this research using KerasTuner to tune hyperparameters and find the optimal network architecture has been successfully applied in ongoing studies to improve the precision of timing computations in the CMS-PPS subsystem at the LHC. 

The Large Hadron Collider (LHC) restarted in 2022 and is producing data in a format similar to that discussed in this article. The SAMPIC readout board saves the full waveform in the raw data stream, which can be subject to further analysis using the method presented in this manuscript. Although the CFD is still commonly used for timing measurements, it may be replaced by neural networks. Lack of the MCP in the LHC setup poses a problem in terms of the reference data acquisition, but the work on finding another way is ongoing.

The inference from the neural network has a very low demand for the CPU time (order of 10 milliseconds), making it well-suited for online processing, such as in the high-level trigger reconstruction chain. The data can be processed in batches, enabling efficient parallel processing of a large number of events.

\section*{Credits}

The author list is formed alphabetically, except for the two main authors listed at the beginning.

\textbf{Mateusz Kocot}, \textbf{Krzysztof Misan}: Methodology, Software, Investigation, Writing - Original Draft
\textbf{Valentina Avati}: Validation, Resources, Writing - Review \& Editing, Supervision, Project administration
\textbf{Edoardo Bossini}: Validation, Resources, Data Curation, Writing - Review \& Editing, Supervision
\textbf{Leszek Grzanka}: Resources, Writing - Review \& Editing, Supervision, Project administration
\textbf{Nicola Minafra}: Conceptualisation, Methodology, Validation, Resources, Writing - Review \& Editing, Supervision

\section*{Acknowledgements}

The project was partially funded by the Polish Ministry of Education and Science, project 2022/WK/14. This research was supported in part by PL-Grid Infrastructure. The numerical experiment was possible through computing allocation on the Ares system at ACC Cyfronet AGH under the grant plgccbmc11.

\printbibliography[heading=bibintoc]

\end{document}